\documentclass[aps,pre,twocolumn,floatfix,showpacs,superscriptaddress,longbibliography,nofootinbib]{revtex4-2}

\usepackage{amssymb}
\usepackage{amsmath}
\usepackage{bm}
\usepackage{graphicx}
\usepackage{hyperref}
\usepackage{xcolor}
\usepackage{soul}

\usepackage[utf8]{inputenc}

\begin{document}

\title{Turbulence to order transitions in activity patterned active nematics}

\author{Cody D. Schimming}
\email[]{cschimm2@jh.edu}
\affiliation{Department of Physics and Astronomy, Johns Hopkins University, Baltimore, Maryland, 21218, USA}
\affiliation{Theoretical Division and Center for Nonlinear Studies, Los Alamos National Laboratory, Los Alamos, New Mexico, 87545, USA}

\author{C. J. O. Reichhardt}
\affiliation{Theoretical Division and Center for Nonlinear Studies, Los Alamos National Laboratory, Los Alamos, New Mexico, 87545, USA}

\author{C. Reichhardt}
\affiliation{Theoretical Division and Center for Nonlinear Studies, Los Alamos National Laboratory, Los Alamos, New Mexico, 87545, USA}

\begin{abstract}
We numerically study two-dimensional active nematics with periodic activity patterning. 
For stripes of activity, we observe
a transition from two-dimensional to one-dimensional active turbulence 
as the maximum active force and distance between activity stripes increases,
followed by
a transition to stable vortices 
ordered antiferromagnetically along the stripes and
ferromagnetically transverse to the stripes.
By comparing to a triangular lattice of activity circles, we find that transitions to
two-dimensional active turbulence emerge from interplays between the active length scale and activity density, independent of the patterning geometry.
The vortex ordering, on the other hand, is highly sensitive to patterning geometry, which we show by comparing the activity stripes to columns of activity circles, where the vortex
ordering is lost.
Our results provide a mechanism for inducing
non-equilibrium phase transitions in active nematics
using activity inhomogeneity, which can be further exploited to create
activity patterned ordered phases.
\end{abstract}

\maketitle

\section{Introduction}

In non-equilibrium active fluids, energy consumed at
the microscale generates macroscopic flows \cite{marchetti13,Bechinger16}. 
Active nematics, 
where the constituents and forces are also endowed with apolar anisotropy, 
appear in cytoskeletal filament suspensions with molecular motors \cite{Sanchez12,DeCamp15,kumar18,doo18}, motile bacteria dispersed in passive nematic liquid crystals \cite{zhou14b,genkin17}, cellular tissues \cite{duclos14,saw17}, colonies of elongated soil bacteria \cite{copenhagen21}, and acoustically actuated lyotropic chromonic liquid crystals \cite{sokolov24}.
In large, unconfined active nematics, the system exhibits ``active turbulence,'' where the flows are chaotic and topological defect pairs spontaneously unbind \cite{giomi13,thampi14b,shankar18,shankar19,Tan19,carenza20}.
There has been growing interest in developing methods to control
the flows and phase behavior
of active nematics by changing the substrate friction or external boundaries 
\cite{thampi14,wu17,opathalage19,hardouin19,hardouin20,thijssen20,thijssen20b,thijssen21,figueroa22,Mori23,ronning23,schimming23b,Partovifard24,schimming24,schimming24b,shankar24,velez24,velez24b}. 
A particularly promising approach is activity patterning,
recently proven achievable
using light activated motor proteins dispersed in cytoskeletal filament suspensions \cite{Rzhang21,RZhang22,Lemma24,Nishiyama24}.  
Recent theoretical and numerical studies
indicate that activity gradients can separate and orient topological defects and induce multiple active regimes \cite{Tang21,ronning23,Partovifard24,shankar24,Ghosh24,Ceron24}.

Active-passive interfaces and emulsions of passive droplets immersed in active nematics have been recently studied, revealing interesting couplings between the passive and active phases that induce novel flow states in the system \cite{Guillamat18,Coelho19,coelho20,Coelho22,Negro24,Zhao24}.
In these studies, however, the passive system is typically isotropic, so there is no coupling between the active and inactive regions apart from the interface.
A recent study of spatially varying activity patterning focused only on separated, inactive, isotropic regions with a connected active nematic region \cite{Partovifard24}.
In that study, the inactive regions acted similarly to obstacles in the active nematic \cite{figueroa22,schimming24,schimming24b,velez24,velez24b}, rather than inducing a coupling between the active and passive regions of the nematic.
Here we focus on two-dimensional nematics with active regions separated by inactive, but still nematic regions, which allow for coupling between the active and elastic forces over larger length scales. 
This is particularly amenable to experimental implementation using the light-activated motor protein system \cite{Rzhang21,RZhang22,Lemma24,Nishiyama24}, but also has fundamental importance in understanding the effect of inhomogeneous activity on active systems in general.
In real active fluids, the energy consumed at the microscale, be it chemical or mechanical, is not homogeneously distributed, so understanding the effects of inhomogeneous activity is important for interpreting the results of experiments.

Here we study inhomogeneous activity patterning by introducing periodic active and inactive stripes or circles.
In Sec.~\ref{sec:Methods} we delineate our model and numerical methods, and describe the quantitative measures that we use to analyze our results.
In Sec.~\ref{sec:TurbTransition} we present results for a system of periodic activity stripes and show that such patterning can produce
transitions from two-dimensional (2D)
to one-dimensional (1D) active turbulence or
a stable vortical flow state
as the activity strength and 
distance between stripes are varied.
Surprisingly, we find that the transitions to turbulence
shift to
higher active region densities for higher activity.
We confirm this by comparing to a lattice of active circles, and show that the interplay between the active length scale and characteristic activity patterning length result in the turbulence transition.
In Sec.~\ref{sec:VortexOrdering} we examine the vortical flow state in the activity stripe system and show that the vortices order antiferromagnetically along the active stripes and ferromagnetically transverse to them, with stronger ordering as the stripes get further apart.
We also show that although vortices still appear if the activity stripes are replaced by columns
of activity circles, the ordering of the vortices is lost.
Finally, in Sec.~\ref{sec:Conclusion} we summarize our results and suggest directions for future studies.
Our results provide a new mechanism
for stabilizing ordered vortices in active nematics and 
controlling transitions to and from different active turbulence regimes.
Crucially, we find
that the activity does not need to be homogeneously distributed in order
to produce active turbulence,
and that the density of activity itself can control the phase
behavior of an active nematic system.

\section{Methods} \label{sec:Methods}

\subsection{Active nematic model with spatially varying activity}

We model an active nematic using the Beris-Edwards equations for the two-dimensional tensor order parameter $\mathbf{Q} = S\left[\mathbf{n}\otimes\mathbf{n} - (1/2)\mathbf{I}\right]$, where $S$ is the degree of local order, $\mathbf{n}$ is the local orientation of the nematic (the director), and $\mathbf{I}$ is the $2\times2$ identity matrix, in the presence of a fluid flow $\mathbf{v}$ \cite{deGennes75,beris94}: 
\begin{equation} \label{eqn:QEvoFull}
    \frac{\partial \mathbf{Q}}{\partial t} + \left(\mathbf{v}\cdot \nabla\right)\mathbf{Q} = \mathbf{S} - \frac{1}{\gamma}\left[\frac{\delta F}{\delta \mathbf{Q}}\right]^{TS}
\end{equation}
where $\gamma$ is a rotational viscosity, $[\cdot]^{TS}$ denotes the traceless, symmetric part of a tensor,
\begin{equation}
    F = \int\left(A|\mathbf{Q}|^2 + C|\mathbf{Q}|^4 + L|\nabla \mathbf{Q}|^2\right) \, d\mathbf{r}
\end{equation}
is the two-dimensional Landau-de Gennes free energy with material parameters $A$ and $C$ and elastic constant $L$, and
\begin{multline}
\mathbf{S} = \left(\lambda\mathbf{E} + \bm{\Omega}\right)\left(\mathbf{Q} + \frac{1}{2}\mathbf{I}\right) + \left(\mathbf{Q} + \frac{1}{2}\mathbf{I}\right)\left(\lambda\mathbf{E} - \bm{\Omega}\right) \\ - 2\lambda\left(\mathbf{Q} + \frac{1}{2}\mathbf{I}\right)\left(\nabla \mathbf{v} : \mathbf{Q}\right)
\end{multline}
is a generalized tensor advection with $\mathbf{E}$ the strain rate tensor, $\bm{\Omega}$ the vorticity tensor, and $\lambda$ a dimensionless parameter that characterizes the tendency of the director to align or tumble under shear flows \cite{beris94}.
The flows in the active nematic are generated by an active stress that arises due to inhomogeneity in the nematic texture.
These are modeled with a modified Stokes equation for zero Reynolds number flows \cite{Marenduzzo07b,doo18}:
\begin{equation} \label{eqn:StokesFull}
    \eta \nabla^2\mathbf{v} - \Gamma \mathbf{v} = \nabla p + \nabla \cdot \left(\alpha \mathbf{Q}\right)
\end{equation}
where $\eta$ is the shear viscosity, $\Gamma$ is the coefficient of friction with the substrate, $p$ is the fluid pressure, and $\alpha = \alpha(x,y)$ is the inhomogeneous strength of active forces.
We assume that the fluid is incompressible so that $\nabla \cdot \mathbf{v} = 0$, and thus the fluid pressure is determined from this constraint.
\textcolor{black}{We note that we neglect elastic stresses that may appear in Eq.~\eqref{eqn:StokesFull} as they are higher order in $\mathbf{Q}$ or gradients of $\mathbf{Q}$, and it has been shown that the active stress dominates the elastic stresses when they are included in Eq.~\eqref{eqn:StokesFull} \cite{giomi13,Pismen13,giomi14}.}
\textcolor{black}{We expect the results presented below to hold for models which include elastic stresses, though the quantitative values of critical parameters may differ.}

There are three important length scales intrinsic to the model. 
The nematic correlation length $\xi = \sqrt{\varepsilon L/C}$ is the length scale over which the nematic distorts, where $\varepsilon$ is a dimensionless parameter which controls the size of topological defects. The hydrodynamic theory induces two more length scales: the hydrodynamic screening length $\ell_h = \sqrt{\eta/\Gamma}$ is the scale over which vortices decay due to substrate friction, while the active length $\ell_{\alpha} = \sqrt{L/\alpha}$ is the scale over which active forces perturb the underlying nematic order. 
The active length scale is set by competing elastic and active forces and is related to the average separation of topological defects. 
In finite sized systems, if the active length is larger than the system size, elastic forces dominate and defect nucleation is suppressed. 
On the other hand, if the active length is much smaller than the system size, the system will be characterized by active turbulence \cite{doo17,opathalage19}.

\textcolor{black}{We note that, in this study, we modify Eq.~\eqref{eqn:StokesFull} by \textcolor{black}{first dividing by $\Gamma$, then multiplying the active force term by $(\eta + \xi^2\Gamma)/(\eta + \xi^2 \Gamma)$}. After these manipulations, Eq.~\eqref{eqn:StokesFull} may be written as
\begin{equation} \label{eqn:StokesLH}
    \ell_h^2 \nabla^2 \mathbf{v} - \mathbf{v} = \nabla\left(\frac{p}{\Gamma}\right) + \left(\xi^2 + \ell_h^2\right)\nabla\cdot\left(\frac{\alpha}{\eta + \xi^2\Gamma} \mathbf{Q}\right).    
\end{equation}
We do this so that the parameter $\alpha/(\eta + \xi^2\Gamma)$ is clearly the parameter responsible for the strength of flows \cite{SuppNote24}.
}

\textcolor{black}{We now introduce a scheme to non-dimensionalize the above equations, and reduce the system to dimensionless parameters. 
We normalize all lengths by the nematic correlation length $\xi$ and times by the nematic relaxation time $\tau = \gamma/(\xi^2 C)$.
We may then represent dimensionless space and time by
\begin{equation}
    \mathbf{\tilde{r}} = \frac{\mathbf{r}}{\xi}, \, \tilde{t} = \frac{t}{\tau}.
\end{equation}
Defining
\begin{equation}
    \mathbf{\tilde{v}} = \frac{\tau}{\xi}\mathbf{v}, \, \mathbf{\tilde{S}} = \frac{\mathbf{S}}{\tau}, \, \tilde{F} = \frac{F}{\xi^2 C}, \, \tilde{A} = \frac{A}{C}
\end{equation}
allows us to rewrite Eq.~\eqref{eqn:QEvoFull} in a dimensionless form:
\begin{equation} \label{eqn:QEvoDimLess}
    \frac{\partial\mathbf{Q}}{\partial\tilde{t}} + \left(\mathbf{\tilde{v}}\cdot\tilde{\nabla}\right)\mathbf{Q} = \mathbf{\tilde{S}} - \left[\frac{\delta \tilde{F}}{\delta \mathbf{Q}}\right]^{TS}
\end{equation}
where $\tilde{\nabla}$ denotes a derivative with respect to $\mathbf{\tilde{x}}$ \textcolor{black}{and
\begin{equation}
\tilde{F} = \int \left(\tilde{A} |\mathbf{Q}|^2 + |\mathbf{Q}|^4 + \frac{1}{\varepsilon}|\tilde{\nabla}\mathbf{Q}|^2\right) \, d\tilde{\mathbf{r}}.
\end{equation}
}
Similarly, we define the dimensionless parameters
\begin{equation}
    \tilde{\ell_h} = \frac{\ell_h}{\xi}, \, \tilde{p} = \frac{\tau p}{\xi^2 \Gamma}, \, \tilde{\alpha} = \frac{\tau \alpha}{\eta + \xi^2\Gamma}
\end{equation}
which allow us to rewrite Eq.~\eqref{eqn:StokesLH} as
\begin{equation} \label{eqn:StokesDimLess}
    \tilde{\ell_h}^2\tilde{\nabla}^2\mathbf{\tilde{v}} - \mathbf{\tilde{v}} = \tilde{\nabla}\tilde{p} + \left(1 + \tilde{\ell_h}^2\right)\tilde{\nabla}\cdot \left(\tilde{\alpha} \mathbf{Q}\right).
\end{equation}
We note that under this non-dimensionalization scheme we may also define a dimensionless active length $\tilde{\ell_{\alpha}} = \ell_{\alpha}/\xi$ such that $\tilde{\ell_{\alpha}} \propto 1/\sqrt{\varepsilon \tilde{\alpha}}$.
For the remainder of the paper we omit the tildes and work in dimensionless units.
In particular, all lengths are given in units of the nematic correlation length.
}

\begin{figure}
\centering
    \includegraphics[width = \columnwidth]{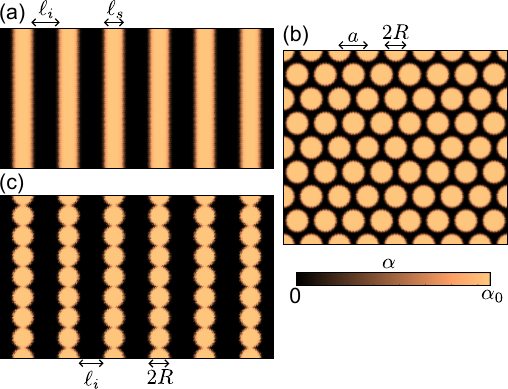}
    \caption{Height maps of activity profiles showing activity level $\alpha$.
    (a) Active stripes with active region width $\ell_s$ and inactive region width $\ell_i$.
    (b) Active circles of radius $R$ arranged in a triangular lattice with lattice constant $a$.
    (c) Active circles of radius $R$ arranged along vertical columns with inactive width $\ell_i$.}
    \label{fig:ActivityProfiles}
\end{figure}

To study activity inhomogeneity, we consider rectangular domains of various sizes with different patterns of activity $\alpha(x,y)$ set to a maximum value of $\alpha = \alpha_0$ and minimum value of $\alpha = 0$.
Figure \ref{fig:ActivityProfiles}(a) shows a system of active stripes oriented along the $y$ direction, where the active (inactive) regions are of length $\ell_s$ ($\ell_i$) \textcolor{black}{relative to the nematic correlation length}.
The equation for the stripe pattern is
\begin{equation} \label{eqn:AlphaStripes}
    \alpha(x,y) = \frac{\alpha_0}{2}\sum_{n=1}^{N=6}\left[1 + \tanh\left(\frac{\ell_s^2/4 - (x - C_n)^2}{\delta^2}\right)\right]
\end{equation}
where $C_n$ give the strip centers which are equally spaced at intervals of $\ell_s + \ell_i$ and $\delta$ is a parameter that controls the width of the interface between active and inactive regions.
\textcolor{black}{We study systems with six activity stripes so that the system size in the $x$ direction is $L_x = 6(\ell_i + \ell_s)$ while we fix the system size in the $y$ direction to be $L_y = 8\ell_h$.}

We also consider active circles of radius $R$ arranged in either a triangular lattice with lattice constant $a$ [Fig.~\ref{fig:ActivityProfiles}(b)] or vertical columns with inactive length spacing $\ell_i$ [Fig.~\ref{fig:ActivityProfiles}(c)].
In either case, the equation for the activity is
\begin{equation} \label{eqn:AlphaCircles}
    \alpha(x,y) = \frac{\alpha_0}{2}\sum_{n=1}^{N}\left[1 + \tanh\left(\frac{R - |\mathbf{r} - \mathbf{C}_n|}{\delta}\right)\right]
\end{equation}
where $\mathbf{r} = (x,y)$ and $\mathbf{C}_n$ is the vector position of the circle centers.
\textcolor{black}{For the lattice of activity circles, the number of activity circles varied with the lattice constant $a$ (or area fraction $\nu$) so that the system size in the $x$ direction $L_x = \sqrt{N} a$ while the system size in the $y$ direction was $L_y = \sqrt{3 N} a/2$.
For the smallest area fractions we used $N = 4$ activity circles, while for the largest area fractions we used $N = 100$, choosing the number that kept $L_x$ as close to $120$ as possible.
For the vertical columns of circles we used a fixed $N = 48$ active circles with system sizes $L_x = 6(\ell_i + 2R)$ and $L_y = 16 R$.
}

\subsection{Numerical Implementation}

To numerically solve Eqs.~\eqref{eqn:QEvoDimLess} and \eqref{eqn:StokesDimLess}, we write $\mathbf{Q}$ in a basis for symmetric, traceless matrices, giving two degrees of freedom $q_1$ and $q_2$.
Equation \eqref{eqn:QEvoDimLess} is then rewritten in terms of $q_1$ and $q_2$. We discretize the equations in space using the finite element Matlab/C++ package FELICITY \cite{walker18} and write Eqs.~\eqref{eqn:QEvoDimLess} and \eqref{eqn:StokesDimLess} in weak form, using the pressure as a Lagrange multiplier for the incompressibility constraint.
We note that spatial derivatives of $\alpha$ also contribute to the flow when $\alpha$ is inhomogeneous.
We compute these from Eqs.~\eqref{eqn:AlphaStripes} and \eqref{eqn:AlphaCircles}.

Given a nematic configuration $\mathbf{Q}$, we first compute the corresponding flow velocity from Eq.~\eqref{eqn:StokesDimLess}. The resulting velocity is then used to solve for the updated $\mathbf{Q}$ from Eq.~\eqref{eqn:QEvoDimLess} using a backwards Euler method with a Newton-Raphson sub-routine to solve the resulting nonlinear equations.
For all simulations we use periodic boundary conditions on the rectangular domain, the size of which is determined by the inactive length $\ell_i$ or lattice constant $a$.
For all simulations we set $A = -2$, $\lambda = 1$, $\varepsilon = 4$, and the Euler time step $\delta t = 0.5$.
We also fix the hydrodynamic screening length and the activity patterning length scales to be commensurate, so $\ell_h = 10$, $\ell_s = 10$, and $R = 10/\sqrt{\pi}$ are not varied in this study.
We do this so the size of the active regions will be roughly the size of a single fluid vortex.
This is to promote comparison with previous studies of active nematics in channel geometries that produced vortices of roughly channel height \cite{doo17,shendruk17}.
The parameters that we vary are the maximum activity $\alpha_0$ and the inactive region lengths $\ell_i$ or $a$.

We initialize the system with a random director configuration and simulate in the range of $300$ to $3000$ time steps depending on how long the system takes to reach a stationary configuration or a dynamical steady state. We define a system to be in a dynamical steady state when the time average of measured quantities does not appreciably change if the time averaging is performed over later times.

\subsection{Measures}

To analyze the behavior of the activity patterned active nematic system, we use several measures involving the spatially varying fields $\mathbf{Q}$ and $\mathbf{v}$.
To understand the distribution of topological defects --- singular points in the director field around which the director winds by half integer multiples of $2\pi$ --- we compute the topological charge density \cite{blow14,schimming22}
\begin{equation}
    D = \frac{1}{2\pi S_N^2}\varepsilon_{k\ell}\varepsilon_{\mu \nu}\partial_k Q_{\mu \alpha} \partial_{\ell} Q_{\nu \alpha}
\end{equation}
where \textcolor{black}{$S_N$ is the value of $S$ in the nematic phase (here $S_N = \sqrt{2}$)}, $\bm{\varepsilon}$ is the two-dimensional fully antisymmetric tensor, $\partial_k \equiv \partial / \partial x_k$, and summation on repeated indices is assumed.
$D$ has the property that it is zero away from defects, positive at the core of $+1/2$ winding defects, and negative at the core of $-1/2$ defects.

We also use the statistics of the vorticity field, $\omega = \left(\nabla \times \mathbf{v}\right)_z$, to analyze the state of a system.
In particular, we make use of the excess kurtosis of the vorticity distribution, defined by
\begin{equation} \label{eqn:Kurtosis}
    K\left[\mathcal{P}(\omega)\right] = \frac{\mu_4}{\sigma^4} - 3
\end{equation}
where $\mu_4$ is the fourth central moment of the vorticity distribution $\mathcal{P}(\omega)$ and $\sigma$ is its standard deviation.
The distribution of vorticities is sampled over times within a stationary configuration or a dynamical steady state, a few examples of which may be seen in Fig.~\ref{fig:KurtosisPhase}(a).
For a Gaussian distribution $K = 0$, hence we use this as a measure of deviation from Gaussian statistics.
In general, larger excess kurtosis indicates larger distribution tails or a sharper distribution peak.
We note that excess kurtosis, and more generally vorticity distribution shape, has been previously used as a measure of active nematic phase behavior \cite{doo17,shendruk17,lemma19,Partovifard24}.

To understand any ordering in the vorticity field, we also measure the correlation function of the vorticity field in either the $x$ and $y$ directions, given by
\begin{align}
    C_x(X,y_0,t) &= \frac{\langle\omega(x + X,y_0,t)\omega(x,y_0,t)\rangle_x}{\langle\omega(x,y_0,t)^2\rangle_x} \label{eqn:CorrX} \\
    C_y(x_0,Y,t) &= \frac{\langle\omega(x_0,y + Y,t)\omega(x_0,y,t)\rangle_y}{\langle\omega(x_0,y,t)^2\rangle_y} \label{eqn:CorrY}
\end{align}
where $x_0$ or $y_0$ are fixed values and $\langle\cdot\rangle_{x,y}$ denotes an average over $x$ or $y$.
Vortex ordering along these directions is identified with peaks in $C_{x,y}$, which we discuss further in Sec.~\ref{sec:VortexOrdering}.

\section{Turbulence Transitions} \label{sec:TurbTransition}

\subsection{Activity Stripes}

\begin{figure*}
\centering
    \includegraphics[width = \textwidth]{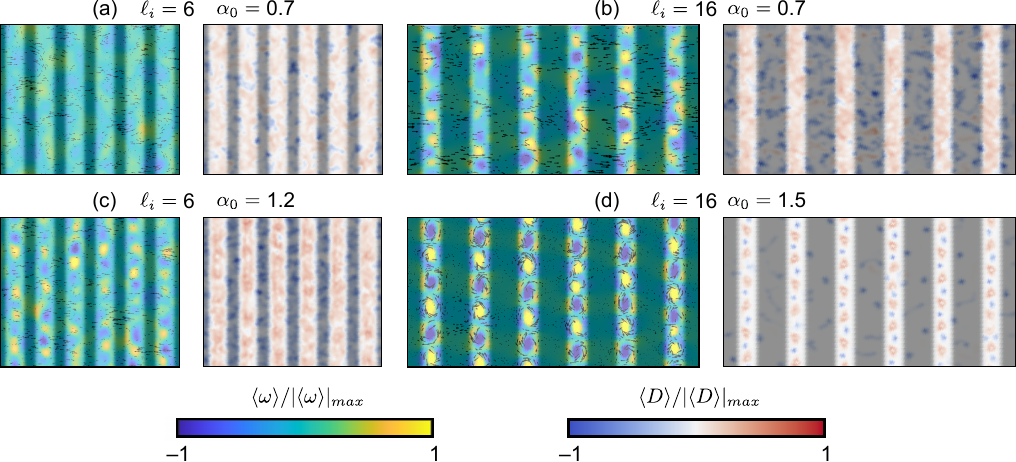}
    \caption{(a-d) Spatial plots of (left) scaled time-averaged vorticity field $\langle \omega \rangle/|\langle \omega \rangle|_{max}$ (color) with time-averaged velocity field $\langle \mathbf{v} \rangle$ (arrows) and (right) scaled time-averaged topological defect density $\langle D \rangle/|\langle D\rangle|_{max}$, where
bright (dark) areas indicate $\alpha=\alpha_0$ ($\alpha=0$). The length of the black arrows indicate the relative magnitude of $\langle \mathbf{v} \rangle$.
     (a) $\ell_i = 6$ and $\alpha_0 = 0.7$.
(b)
$\ell_i = 16$ and $\alpha_0 = 0.7$.
(c)
$\ell_i = 6$ and $\alpha_0 = 1.2$.
(d)
$\ell_i = 16$ and $\alpha_0 = 1.5$.
Animations of the dynamics of each state are available
in the Supplemental Material \cite{SuppNote24}.
}
    \label{fig:AverageVorticity}
\end{figure*}

We first focus on the qualitative transitions observed in the activity stripe system [Fig.~\ref{fig:ActivityProfiles}(a)] upon varying $\ell_i$ and $\alpha_0$.
For small $\ell_i$ and $\alpha_0$, the behavior resembles
the active turbulence that would appear if the activity were homogeneous
\cite{DeCamp15,doo17,doo18,lemma19}. 
We show an example of this in Fig.~\ref{fig:AverageVorticity}(a), where we plot the time-averaged normalized vorticity $\langle \omega \rangle/|\langle\omega\rangle|_{max}$ and velocity $\langle \mathbf{v} \rangle$ fields as well as the time averaged defect density $\langle D \rangle/|\langle D\rangle|_{max}$ for a system with $\ell_i = 6$ and $\alpha_0 = 0.7$.
The average velocity and vorticity are close to zero throughout the domain, with no correlation between the active regions and the weak vortical patterns.
Likewise, the topological defect density is also not correlated with the active regions, unlike in systems with a single activity stripe \cite{shankar24} or active channels \cite{shendruk17}.

When either $\alpha_0$ or $\ell_i$ is increased,
as shown
in Fig.~\ref{fig:AverageVorticity}(b) at $\ell_i=16$ and $\alpha_0=0.7$
and in Fig.~\ref{fig:AverageVorticity}(c) at $\ell_i=6$ and $\alpha_0=1.2$,
%a charge separating behavior arises, more in line with previous observations for active nematics in a single active stripe or confined in a channel \cite{shendruk17,shankar24}.
$\langle \omega\rangle/|\langle \omega \rangle|_{max}$ becomes nonzero in the active regions while
$|\langle \mathbf{v} \rangle|$ is suppressed
in the active regions, \textcolor{black}{which can be seen by the length of the black arrows in Fig.~\ref{fig:AverageVorticity}}, so that steady flows occur only in the inactive
regions between the active stripes.
Additionally, the time-averaged topological defect density separates, and
positive topological defects occupy the active regions while
negative topological defects remain in the inactive regions.
This charge separating behavior is similar to the behavior found in previous observations of active nematics in a single active stripe or confined in a channel \cite{shendruk17,shankar24}.

At large $\alpha_0$, the system develops stable and sometimes static vortices,
as illustrated
in Fig.~\ref{fig:AverageVorticity}(d)
at $\ell_i=16$ and $\alpha_0=1.5$, where
$\langle \mathbf{v} \rangle$ has clear vortical signatures.
\textcolor{black}{By plotting the time averages $\langle \omega \rangle$ and $\langle \mathbf{v}\rangle$ in this state versus those in Figs.~\ref{fig:AverageVorticity}(a,b,c), we see a clear difference in the long-time dynamical behavior of the system.}
The nematic configuration in this regime
consists of two positive defects rotating around one another
near the center of a given vortex,
while negative defects sit in between the vortices.
We discuss this state in more detail in Sec.~\ref{sec:VortexOrdering}.

\begin{figure}
\centering
    \includegraphics[width = \columnwidth]{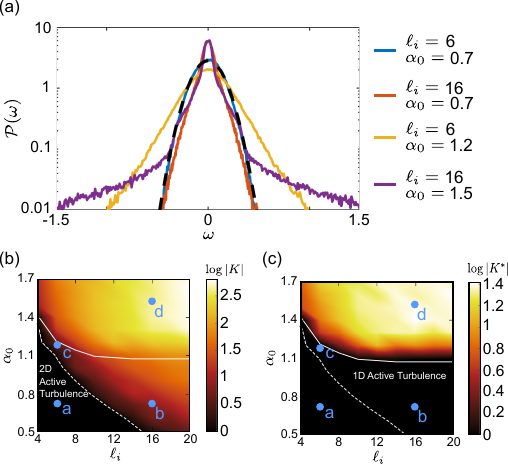}
    \caption{(a) Vorticity distributions, $\mathcal{P}(\omega)$, for simulations with $(\ell_i,\alpha_0) = (6,0.7),\,(16,0.7),\,(6,1.2),\,(16,1.5)$ (the same parameter sets in Fig.~\ref{fig:AverageVorticity}). 
    The dashed black line shows a Gaussian distribution.
    (b) Heat map of the logarithm of the excess kurtosis of the vorticity distribution over the whole domain, $\log|K|$, as a function of $\ell_i$ and $\alpha_0$.
    (c) Heat map of the excess kurtosis of the vorticity distribution measured over active regions only, $\log|K^*|$, as a function of $\ell_i$ and $\alpha_0$.
    Dashed line: points from panel (b) where $\log|K| = 0$.
    Solid line: points from panel (c) where $\log|K^*| = 0$.
    Letters correspond to parameter sets shown in Fig.~\ref{fig:AverageVorticity}.}
    \label{fig:KurtosisPhase}
\end{figure}

To quantify the apparent transitions to and from active turbulence, we examine the vorticity distribution $\mathcal{P}(\omega)$ across the transitions.
In Fig.~\ref{fig:KurtosisPhase}(a) we plot examples of $\mathcal{P}(\omega)$ for the same parameter sets shown in Fig.~\ref{fig:AverageVorticity}.
Only the simulation that qualitatively resembles two-dimensional active turbulence --- $\ell_i = 6$ and $\alpha_0 = 0.7$ --- is characterized by a Gaussian vorticity distribution, shown by the black dashed line in Fig.~\ref{fig:AverageVorticity}(a).
The other distributions are sharply peaked around $\omega = 0$ and/or have long tails.
We further characterize the distribution by the excess kurtosis of the distribution $K$, defined in Eq.~\eqref{eqn:Kurtosis}.
In Fig.~\ref{fig:KurtosisPhase}(b) we plot a heat map of $\log|K|$
measured over the entire spatial domain
as a function of $\alpha_0$ and $\ell_i$.
The dashed line indicates points where $\log|K|=0$, and we define
the
$\log|K| < 0$
region below this line
to be in 2D active turbulence.
We demarcate $K = 1$ as the line for the 2D active turbulence transition because below this line the values of $K$ for each simulation do not vary much and tend to be close to $K=0$ which indicates a perfect Gaussian distribution.
\textcolor{black}{Additionally, we find that for systems characterized by vorticity distributions with $K > 1$, the excess kurtosis increases with increasing $\alpha_0$ or $\ell_i$, while for systems with $K < 1$ the small fluctuations in $K$ are not correlated with $\alpha_0$ or $\ell_i$.}

Above this transition, we observe that the qualitative behavior is similar to active systems confined to a channel or single activity stripe \cite{doo17,shendruk17,shankar24}.
To probe this, we perform a measurement of
$K[\mathcal{P}(\omega | \alpha > 0.5\alpha_0)]\equiv K^*$, in which we restrict
the domain to only the active portions of the sample and exclude
the inactive areas. 
We plot the
resulting heat map of $\log|K^*|$ as a function of
$\alpha_0$ versus $\ell_i$ in
Fig.~\ref{fig:AverageVorticity}(c).
The solid line indicates points for which $\log|K^*|=0$.
In the window between the solid and dashed lines, where the active regions
produce a turbulent signature but the inactive regions do not,
1D active turbulence
has emerged along each active stripe,
but turbulent behavior is suppressed in the regions between the stripes.
Thus, the two transitions in the system are between 2D and
1D active turbulence, and between 1D turbulence
and a vortex state.

\subsection{Activity Circles} 

\begin{figure}
\centering
    \includegraphics[width = \columnwidth]{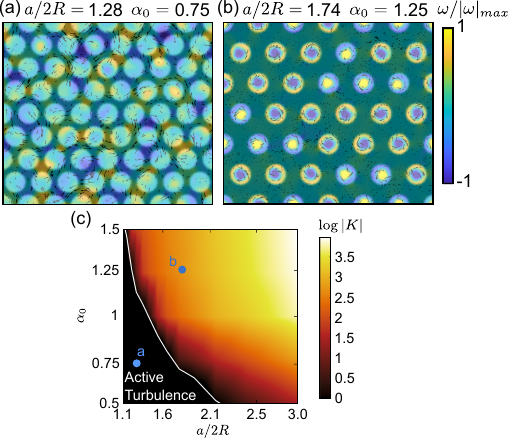}
    \caption{(a,b) Time snapshots of $\omega/|\omega|_{max}$ (color) and
      ${\bf v}$ (arrows) for a triangular lattice of active circles (bright spots).
      (a) $a/2R=1.28$ and $\alpha_0=0.75$.
      (b) $a/2R=1.74$ and $\alpha_0=1.25$.
      Animations of the dynamics of each state are available in the
      Supplemental Material \cite{SuppNote24}.
      (c) Heat map of
      $\log|K|$ as a function of $\alpha_0$ vs $a/2R$.
      Line: points where $\log|K| = 0$. Letters: locations at which images in
      panels (a,b) were obtained.}
    \label{fig:CircleLattice}
\end{figure}

To further probe
the turbulent to vortex transition,
we simulate a system in which the activity is confined to circles
that are arranged in a triangular lattice,
as shown in Fig.~\ref{fig:ActivityProfiles}(b).
The area fraction of the active regions
is given by $\nu = (\pi/2\sqrt{3})(2R/a)^2$,
where $R$ is the circle radius and $a$ is the lattice constant.
We vary $a/2R$ to give values of $\nu$ in the range
$\nu=0.1$ to $\nu=0.7$.
Even without the 1D stripe arrangement of activity,
we observe a transition from active turbulence to a
stable vortex state when $\alpha_0$ and $a/2R$ are increased.
In Fig.~\ref{fig:CircleLattice}(a) we plot a time snapshot of
the vorticity and velocity fields for a simulation
with $a/2R = 1.28$ and $\alpha_0 = 0.75$.
Here we observe
active turbulence in which the locations of the vortices
are not correlated with the locations of the activity spots.
When we increase $\alpha_0$ and $a/2R$, as shown in Fig.~\ref{fig:CircleLattice}(b) for $\alpha_0 = 1.25$ and $a/2R = 1.74$,
individual vortices
become localized in each active region. 
In Fig.~\ref{fig:CircleLattice}(c),
we plot a heat map of $\log|K|$ measured over the entire domain
as a function of $\alpha_0$ versus $a/2R$.
Below the line marking points at which $\log|K| = 0$
the system exhibits active turbulence,
and above the line the system relaxes to a configuration with a
stable vortex
at each active site.
The turbulent to vortex transition has a similar form to what we
observed in the active stripe sample from
Fig.~\ref{fig:AverageVorticity}(e,f), but the 1D active
turbulent state is lost
due to the geometry of the active circles.

\begin{figure}
\centering
    \includegraphics[width = \columnwidth]{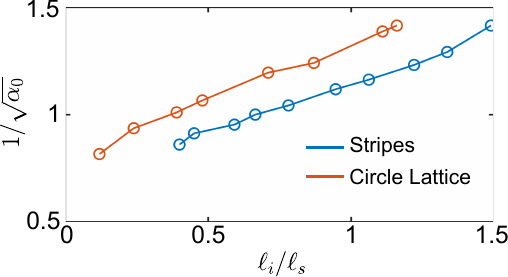}
    \caption{2D active turbulence transition curves for activity stripes and activity circle lattice plotted against $1/\sqrt{\alpha_0}$ versus $\ell_i/\ell_s$. For the lattice of activity circles we have made the identification $\ell_i/\ell_s = (a - 2R)/2R$.}
    \label{fig:TransitionCurves}
\end{figure}

It is surprising that,
for both the active stripes and the active circles,
the transition to the turbulent state shifts to
smaller spacings between the active regions (and a greater active
area fraction) when the activity level
$\alpha_0$ is increased since,
in confined systems with homogeneous activity, an increase in the activity
typically produces an increased amount of 
topological defect unbinding and
a more rapid transition to turbulence
\cite{DeCamp15,doo17,doo18}.
In the systems
with inhomogeneous activity studied here,
we find that when
$\ell_{\alpha}
\gtrsim \ell_i$,
the system exhibits active turbulence,
while when $\ell_{\alpha}
\lesssim \ell_i$,
localized vortical structures appear.
We show this in Fig.~\ref{fig:TransitionCurves} by plotting the 2D active turbulence transition curves from both the active stripe and circle systems in Figs.~\ref{fig:KurtosisPhase}(b) and \ref{fig:CircleLattice}(c) against axes $1/\sqrt{\alpha_0}$ and $\ell_i/\ell_s$.
In the figure, we have identified $\ell_i/\ell_s = (a - 2R)/2R$ for the activity circles.
For both systems, the transitions depend linearly on $1/\sqrt{\alpha_0} \propto \ell_{\alpha}$ as functions of $\ell_i/\ell_s$, indicating that the transitions are given by a direct interplay between the two length scales.

The dependence of the transition on activity density has implications even for systems that do not have explicitly patterned activity, since in any experimental system the active forces will not be distributed homogeneously.
Our results indicate that the approximation of
homogeneously distributed activity is
valid in the limit of $\ell_{\alpha} >> \ell_{i}$,
but outside this limit, the effect of inhomogeneous activity
may disrupt the active turbulent phase.

\section{Ordering of Vortices} \label{sec:VortexOrdering}

\subsection{Activity Stripes}

\begin{figure}
\centering
    \includegraphics[width = \columnwidth]{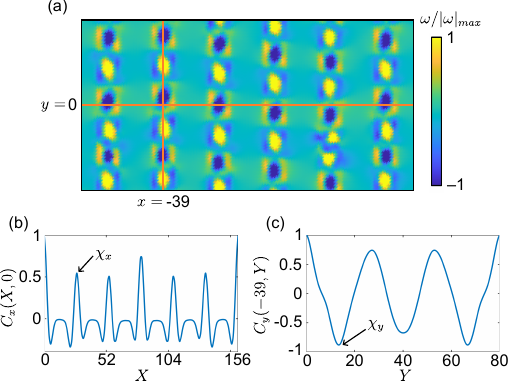}
    \caption{Graphic example of the calculation of $\chi_x$ and $\chi_y$. (a) Frame of the vorticity field for a simulation with $\ell_i = 16$ and $\alpha_0 = 1.5$. The orange lines indicate $x_0 = -39$ and $y_0 = 0$. (b) Vorticity autocorrelation function along the $x$ direction $C_x(X,y_0)$. The value of the first peak gives $\chi_x$. (c) $C_y(x_0,Y)$. The value of the first peak gives $\chi_y$.}
    \label{fig:CorrelationCalculation}
\end{figure}

We now investigate the high activity region of the phase diagram of activity stripes, characterized by well-defined, slowly evolving vortices.
A more careful inspection of
this vortical flow state
reveals strong antiferromagnetic ordering of the vortices
along the $y$ direction
and
a tendency for ferromagnetic ordering along the $x$ direction.
Using the one-dimensional correlation functions defined in Eqs.~\eqref{eqn:CorrX} and \eqref{eqn:CorrY}, $C_x(X,y_0,t)$ and $C_y(x_0,Y,t)$, we may quantify this ordering by
defining $\chi_x(t) = \langle C_x(\ell_s + \ell_i,y_0,t) \rangle_{y_0}$ and $\chi_y(t) = \langle C_y(x_0,\ell_h,t) \rangle_{x_0}$ so that $\chi_x$
($\chi_y$) measures the nearest neighbor correlation in the $x$ ($y$)
direction.
Here, $\chi_{x,y} \in [-1,1]$, with positive (negative) values indicating
ferromagnetic (antiferromagnetic) order and larger magnitudes
indicating stronger order.
We show an example of the calculation of $\chi_x$ and $\chi_y$ for a frame from a simulation with $\ell_i = 16$ and $\alpha_0 = 1.5$
in Fig.~\ref{fig:CorrelationCalculation}.
We select $y_0 = 0$ and $x_0 = -1.5(\ell_i + \ell_s) = -39$, as illustrated in Fig.~\ref{fig:CorrelationCalculation}(a).
Figures \ref{fig:CorrelationCalculation}(b,c) show the corresponding $C_{x,y}$ versus $X$ and $Y$.
For $C_x$ the peaks of the correlation function naturally occur at multiples of $\ell_s + \ell_i$ since the centers of the activity regions are separated by this distance, while the peaks of $C_y$ naturally occur at multiples of $\ell_h$ since this is the characteristic size of vortices.
Thus, the first peak for each of these functions measures the nearest neighbor correlations [note that $C_y$ in Fig.~\ref{fig:CorrelationCalculation}(c) alternates between negative and positive peaks because of the antiferromagnetic order].
We average over multiple values of $y_0$ and $x_0$ to get the values of $\chi_x$ and $\chi_y$ presented below. 

\begin{figure}
\centering
    \includegraphics[width = \columnwidth]{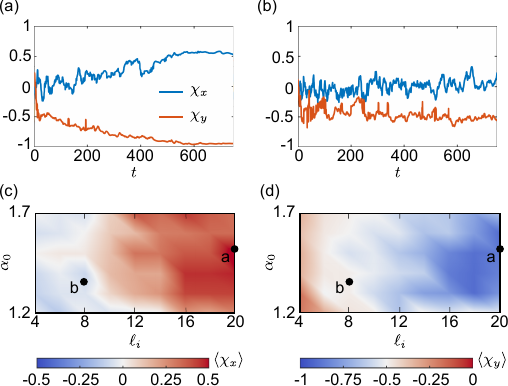}
    \caption{(a, b) Vortex correlation parameters $\chi_x$ and $\chi_y$ versus time. (a) $\ell_i = 20$ and $\alpha_0 = 1.5$.
      (b)
      $\ell_i = 8$ and $\alpha_0 = 1.4$.
    (c) Heat map of the time-averaged $\langle \chi_x \rangle$ as a function of $\alpha_0$ vs $\ell_i$.
      (d) Heat map of the time-averaged $\langle \chi_y \rangle$ as a function of $\alpha_0$ vs $\ell_i$.
    Letters: locations at which curves in panels (a, b) were obtained.}
    \label{fig:VorticityCorrelations}
\end{figure}

We compute $\chi_{x,y}$ for $\alpha_0 > 1.2$,
where $\log|K|$ is large and positive.
We note that below $\alpha_0 < 1.2$ the system is in either 1D or 2D active turbulence, and the vortices are not stable enough to yield meaningful measurements of $\chi_{x,y}$.
We plot $\chi_x(t)$ and $\chi_y(t)$ for a system with $\ell_i = 20$ and $\alpha_0 = 1.5$ in Fig.~\ref{fig:VorticityCorrelations}(a). 
For long times, after the system reaches a stationary state, we find
$\chi_x \approx +0.5$
and $\chi_y \approx -1$,
indicating moderate ferromagnetic order along the $x$ direction
and strong antiferromagnetic order along the $y$ direction.
For smaller $\ell_i$,
$\chi_x$ and $\chi_y$ fluctuate more over time, as
shown in Fig.~\ref{fig:VorticityCorrelations}(b)
at $\ell_i = 8$ and $\alpha_0 = 1.4$.
Here, $\chi_x$ fluctuates around zero but
$\chi_y$ fluctuates around $\chi_y=-0.5$, indicating a lack of ordering
along the $x$ direction and moderate antiferromagnetic order
along the $y$ direction.
In Figs.~\ref{fig:VorticityCorrelations}(c,d) we plot
heat maps of the time averaged
$\langle \chi_x \rangle$ and $\langle \chi_y \rangle$,
respectively, as a function of $\alpha_0$ versus $\ell_i$.
As $\ell_i$ increases,
$\langle \chi_x \rangle$ grows from roughly zero to
$\approx 0.6$, indicating the emergence of
ferromagnetic order transverse to the activity stripes.
In contrast,
$\langle \chi_y \rangle$ is almost always negative,
and reaches values very close to $-1$
when $\ell_i > 14$,
indicating almost perfect, stable antiferromagnetic order
along the activity stripes.

Antiferromagnetic vortex ordering has been observed for active nematics confined to channels, as well as in two-dimensional systems with increasing substrate friction or obstacle confinement \cite{doo17,shendruk17,hardouin19,thijssen20b,schimming24}.
The unbinding of topological defects helps promote antiferromagnetic order when vortices are stable, since the negative defects
can sit between the vortices and mediate a rotation of the nematic director
as the chirality of the positive defects at the center of each
vortex alternates.
We can see this in the time averaged plots of $D$ in Fig.~\ref{fig:AverageVorticity}(d) where the positive defects stay near the center of stable vortices, while single negative defects sit between them.
\textcolor{black}{We note that even though positive defects are typically motile in active nematics, they may form bound states of co-rotation at high activities \cite{Vafa22,Schimming25}.}
\textcolor{black}{In the supplemental movies for simulations in this regime \cite{SuppNote24}, the positive defects sometimes appear as defects with $+1$ winding at the center of the fluid vortices.
This is likely due to the finite resolution of the simulation mesh, since we see in Fig~\ref{fig:AverageVorticity}(d) that the average defect density forms a ring around the center of the vorticies, rather than being concentrated at the center, which indicates that two defects are co-rotating in this region.}

The ferromagnetic order transverse to the active regions is of great interest
since, to our knowledge, it has only been reported numerically for active nematics interacting with periodic arrays of obstacles \cite{schimming24},
while vortex synchronization has only recently been reported in acoustically actuated active nematics \cite{sokolov24}.
For well separated active stripes, the only interaction that couples the stripes is the elastic interaction of the nematic, which has been shown to promote ferromagnetic order of vortices in obstacle laden systems \cite{schimming24}. 
We therefore conclude that the ferromagnetic order transverse to the stripes is driven by elastic interactions, while the antiferromagnetic order along the stripes is driven by active forces.

Interestingly, we find that separation of active regions is necessary for strong vortex ordering.
When the stripes are closer together, flow generation between the stripes disrupts the vortex order,
as illustrated in the Supplemental Movie 7 \cite{SuppNote24},
leading to the fluctuations seen in Fig.~\ref{fig:VorticityCorrelations}(b) and bringing $\langle \chi_{x,y}\rangle$ closer to zero.
These results are more in line with with earlier numerical results on activity patterns with no gaps in the activity \cite{Partovifard24}.

\subsection{Comparison to columns of active circles}

\begin{figure}
\centering
    \includegraphics[width = \columnwidth]{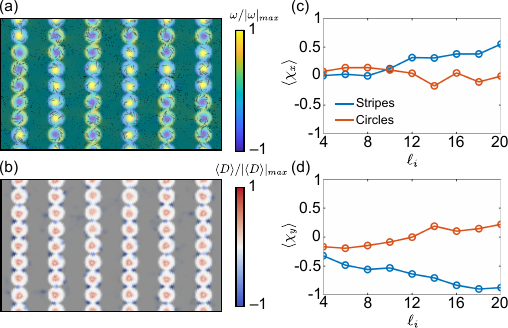}
    \caption{(a) Time snapshot of $\omega/|\omega_{max}|$ (color)
      and $\langle {\bf v}\rangle$ (arrows)
      for columns of activity circles (bright spots) with
      $\ell_i = 14$ and $\alpha_0 = 1.5$.
      (b) The corresponding time-averaged topological defect density.
      (c) $\langle\chi_x\rangle$ and (d) $\langle\chi_y\rangle$ vs
      $\ell_i$ for systems of activity stripes (blue) and circle columns (red)
      with $\alpha_0 = 1.5$.}
    \label{fig:CircleStripes}
\end{figure}

Finally, we investigate the impact of the 1D degree of freedom along the stripes on the vortex ordering.
While the antiferromagnetic order develops quickly along the active stripes, the ferromagnetic order takes much longer to develop.
Qualitatively, the vortices slide into place over time in order to develop
ferromagnetic order along the $x$ direction,
rather than nucleating in ordered positions initially.
We thus suspect that the 1D degree of freedom along the $y$ direction afforded by the stripes is important for observing ordering along the $x$ direction.
To test this, we remove the 1D degree of freedom by simulating columns of active circles [Fig.~\ref{fig:ActivityProfiles}(c)] for $\alpha_0 = 1.5$ at varied $\ell_i$.
As shown in Fig.~\ref{fig:CircleStripes}(a),
single vortices develop in the active regions, similar to both the stripe and circle lattice systems at this activity.
We also show in
Fig.~\ref{fig:CircleStripes}(b) that
positive topological defects rotate in the center of the vortices while negative defects sit between active regions. 

Although the vortex stabilizing behavior in the active circle columns is similar to
that of the active stripe system, the vortex ordering is different.
In Figs.~\ref{fig:CircleStripes}(c) and (d) we plot $\langle \chi_x \rangle$ and $\langle \chi_y \rangle$ versus $\ell_i$ for the stripe and circle column systems.
The measures diverge
for large $\ell_i$,
with
$\langle \chi_x \rangle \approx 0$ and $\langle \chi_y \rangle \gtrsim 0$ for the circle columns, but both measures reaching nonzero values for the stripe system.
We note that, for the active circle columns, the averages $\langle \chi_{x,y} \rangle$ are averaged over time and three separate realizations with different random initial conditions. 
The vortices develop quickly due to the large active forces in the active regions, but in the circle column system,
they remain frozen in place and do not reorder along $x$. 
An example of the dynamics is shown in Supplementary Movie 8.

The measures of $\chi_x$ in Fig.~\ref{fig:CircleStripes}(c) confirm our hypothesis that the 1D degree of freedom is important for observing ferromagnetic ordering along the transverse direction to the stripes.
Interestingly, even though the vortices would naturally adopt an antiferromagnetic
ordering along $y$ in a 1D channel, as in the stripe system, we observe no ordering, or even a slight ferromagnetic ordering when the columns are well separated in the circle column system.
As we noted for the activity stripes, antiferromagnetic ordering requires an odd number of negative topological defects
to sit between pairs of vortices to mediate the change in chirality.
As shown in Fig.~\ref{fig:CircleStripes}(b), in the circle column system,
the divots between adjacent circles permit
two defects to sit between vortices
along the edge of the inactive region.
We find that when one defect is present, the chirality of neighboring
vortices changes,
but when two defects are present, the chirality remains
the same [compare Figs.~\ref{fig:CircleStripes}(a) and \ref{fig:CircleStripes}(b)].
This result underscores
the pivotal role that activity geometry plays in determining the
ordering of the system, as well as the interplay between vortex patterning and topological defect dynamics in active nematic systems.

\section{Conclusion} \label{sec:Conclusion}

We have studied the effects of periodic, inhomogeneous activity patterning on active nematics.
We observe two non-equilibrium transitions for a system of activity stripes: a transition from 2D to 1D active turbulence, and a transition from 1D active turbulence to a well-defined vortical flow state that can exhibit ferromagnetic and antiferromagnetic vortex ordering. 
When we compare the activity stripes to a triangular array of activity circles, the turbulence transitions depend on the interplay between the active length scale and the density of active regions in similar manners, indicating an independence of activity patterning geometry on the transition to turbulence.
On the other hand, we show that vortex ordering depends strongly on patterning geometry by comparing the activity stripes to columns of activity circles, which destroy the observed antiferromagentic and ferromagnetic vortex order.
Our results provide a method for inducing and controlling non-equilibrium phase transitions in active nematics and for producing customized activity patterned ordered states.

It has been recently shown that activity patterning is an experimental possibility \cite{Rzhang21,RZhang22,Lemma24,Nishiyama24}, thus our results should be readily amenable to experimental verification.
An interesting direction for future studies would be to understand the additional role of the hydrodynamic screening length, which we kept fixed in this study.
It is possible that additional non-equilibrium phases may be observed when this parameter is varied and made commensurate with other system length scales.
\textcolor{black}{Additionally, an alternative approach to introduce activity heterogeneity would be to consider a spatially varying substrate friction, which was recently explored experimentally \cite{thijssen21}. 
In the model studied here, this would cause both parameters in Eq.~\eqref{eqn:StokesDimLess} to be spatially varying, and could lead to interesting interplays between the screening length and activity.}
Another direction would be to study the effect of temporally evolving activity fields and how the structure of turbulence or order is preserved or destroyed by motion of the activity fields.
Recent work has shown that there is an optimal ``surfing'' speed for a single stripe of activity \cite{shankar24}, and it would be interesting to study how these results change for multiple stripes or circles of activity.
Finally, while our work gives some insight into the experimentally realistic case of inhomogeneous activity distributions, more insight may be gleaned by studying random activity patterning, as opposed to the periodic patterning considered here.

\begin{acknowledgments}
We gratefully acknowledge the support of the U.S. Department of Energy through the LANL/LDRD program for this work.
This work was supported by the U.S. Department of Energy through the Los Alamos National Laboratory.
Los Alamos National Laboratory is operated by Triad National Security, LLC, for the National Nuclear Security Administration of the U.S. Department of Energy (Contract No. 89233218CNA000001). 
CDS also gratefully acknowledges support through the William H. Miller III fellowship from the Department of Physics and Astronomy at Johns Hopkins University.
\end{acknowledgments}

\bibliography{LC}

\end{document}

\title{Supplemental Material for ``Turbulence to order transitions in activity patterned active nematics''}

\author{Cody D. Schimming}
\affiliation{Department of Physics and Astronomy, Johns Hopkins University, Baltimore, Maryland, 21218, USA}
\affiliation{Theoretical Division and Center for Nonlinear Studies, Los Alamos National Laboratory, Los Alamos, New Mexico, 87545, USA}

\author{C. J. O. Reichhardt}
\affiliation{Theoretical Division and Center for Nonlinear Studies, Los Alamos National Laboratory, Los Alamos, New Mexico, 87545, USA}

\author{C. Reichhardt}
\affiliation{Theoretical Division and Center for Nonlinear Studies, Los Alamos National Laboratory, Los Alamos, New Mexico, 87545, USA}

\maketitle

\section*{Derivation of Equation (5)}
The equation for the velocity field that we solve is slightly modified from the traditional Stokes equation. To get to the modified version, we start with Eq.~(4) in the main text:
\begin{equation*}
    \eta\nabla^2\mathbf{v} - \Gamma\mathbf{v} = \nabla p + \nabla\cdot\left(\alpha \mathbf{Q}\right).
\end{equation*}
We then divide the entire equation by $\Gamma$, which yields
\begin{equation*}
    \ell_h^2 \nabla^2\mathbf{v} - \mathbf{v} = \nabla \left(\frac{p}{\Gamma}\right) + \nabla\cdot\left(\frac{\alpha}{\Gamma} \mathbf{Q}\right)
\end{equation*}
where we have used the definition $\ell_h^2 = \eta / \Gamma$. We then multiply the active force term by $(\eta + \xi^2 \Gamma)/(\eta + \xi^2 \Gamma)$ so that the equation reads
\begin{equation*}
    \ell_h^2 \nabla^2 \mathbf{v} - \mathbf{v} = \nabla \left(\frac{p}{\Gamma}\right) + \left(\frac{\eta + \xi^2 \Gamma}{\Gamma}\right)\nabla\cdot\left(\frac{\alpha}{\eta + \xi^2\Gamma}\mathbf{Q}\right).
\end{equation*}
Simplifying the coefficient on the active force term gives Eq.~(5) in the main text.

Our motivation for renormalizing the parameters as such is that we are able to treat the screening length as independent from the strength of flows that are produced by activity. Consider Eq.~(10) in the main text which gives the dimensionless Stokes equation that we solve. In the wet limit, $\tilde{\ell}_h^2 >> 1$, this equation simplifies to
\begin{equation*}
    \tilde{\nabla}^2 \mathbf{\tilde{v}} = \tilde{\nabla} \left(\frac{\tilde{p}}{\tilde{\ell}_h^2}\right) + \tilde{\nabla}\cdot\left(\tilde{\alpha}\mathbf{Q}\right),
\end{equation*}
while in the dry limit, $\tilde{\ell}_h^2 << 1$, the equation simplifies to
\begin{equation*}
    -\mathbf{\tilde{v}} = \tilde{\nabla}\tilde{p} + \tilde{\nabla}\cdot\left(\tilde{\alpha}\mathbf{Q}\right).
\end{equation*}
In both limits, the magnitude of flows is set by the parameter $\tilde{\alpha}$. This approach differs slightly from previous studies that do not renormalize the parameters and observed arrested flows in the dry limit \cite{thampi14}.

\section*{Supplemental Movie Captions}
\begin{itemize}
    \item Supplemental Movie 1: (Left) Nematic configuration where the color is the scalar order parameter $S$, the white lines show the director $\mathbf{n}$, blue circles label $+1/2$ winding defects, and purple triangles label $-1/2$ winding defects, and (right) vorticity (color) and velocity (black arrows) fields and $\pm1/2$ defects (orange circles and purple triangles) for an activity stripe simulation with stripes $\ell_i = 6$ and $\alpha_0 = 0.7$. The bright regions indicate where $\alpha = \alpha_0$ and dark regions indicate where $\alpha = 0$.
    \item Supplemental Movie 2: Nematic configuration and vorticity and velocity fields for an activity stripe simulation with $\ell_i = 16$ and $\alpha_0 = 0.7$.
    \item Supplemental Movie 3: Nematic configuration and vorticity and velocity fields for an activity stripe simulation with $\ell_i = 6$ and $\alpha_0 = 1.2$.
    \item Supplemental Movie 4: Nematic configuration and vorticity and velocity fields for an activity stripe simulation with $\ell_i = 16$ and $\alpha_0 = 1.5$.
    \item Supplemental Movie 5: Nematic configuration and vorticity and velocity fields for an activity circle triangular lattice simulation with $a/2R = 1.28$ and $\alpha_0 = 0.75$.
    \item Supplemental Movie 6: Nematic configuration and vorticity and velocity fields for an activity circle triangular lattice simulation with $a/2R = 1.74$ and $\alpha_0 = 1.25$.
    \item Supplemental Movie 7: Nematic configuration and vorticity and velocity fields for an activity stripe simulation with $\ell_i = 8$ and $\alpha_0 = 1.4$.
    \item Supplemental Movie 8: Nematic configuration and vorticity and velocity fields for an activity circle vertical column simulation with $\ell_i = 14$ and $\alpha_0 = 1.5$.
\end{itemize}

\bibliography{LC}